\documentclass[11pt]{article}   	
\usepackage{geometry}                		
\usepackage{graphicx}				
\usepackage{amssymb}

\usepackage[superscript,biblabel]{cite}

\usepackage{hyperref}
\usepackage{multicol}

\title{{\bf Dependence of the flux of gamma-ray pulsars on distance: A nonparametric analysis of the data in the second {\it FERMI} catalogue}}
\author{Houshang Ardavan\footnote{ardavan@ast.cam.ac.uk}\\
Institute of Astronomy, University of Cambridge,\\
Madingley Road, Cambridge CB3 0HA, UK}
\date{}							
\begin{document}
\maketitle

\begin{abstract}
Pulsars' gamma-ray luminosities are thought to exceed their radio luminosities by several orders of magnitude: a notion that is based on the assumption that the decay with distance of the flux of gamma-ray pulses obeys the inverse-square law as does that of the flux of radio pulses.  However, results of testing the hypothesis of independence of luminosities and distances of gamma-ray pulsars by means of the Efron--Petrosian statistic imply that the observational data in the second {\it FERMI} catalogue are consistent with the dependence $S\propto D^{-3/2}$ of the flux densities $S$ of these pulsars on their distances $D$ at substantially higher levels of significance than they are with the dependence $S\propto D^{-2}$.  This is not incompatible with the requirements of the conservation of energy because the radiation process described in Ardavan [{\it Mon.\ Not.\ R.\ Astron.\ Soc.}, {\bf 507}, 4530--4563 (2021)], by which the superluminally moving current sheet in the magnetosphere of a neutron star is shown to generate the observed gamma-ray pulses, is intrinsically transient: the difference in the fluxes of power across any two spheres centred on the star is balanced by the change with time of the energy contained inside the shell bounded by those spheres.  Once the over-estimation of their values is rectified, the luminosities of gamma-ray pulsars turn out to have the same range of values as do the luminosities of radio pulsars.
\end{abstract}
\medskip

\begin{multicols}{2}
\section{Introduction}
\label{sec:introduction}

Flux densities of the pulses that are generated by the superluminally moving current sheet in the magnetosphere of a neutron star diminish with the distance $D$ from the star as $D^{-3/2}$ (rather than $D^{-2}$) in certain latitudinal directions (ref.~\citeonline{Ardavan2021}, Section 5.5).  By virtue of their extremely narrow widths in the time domain, such pulses in addition have broad spectra that peak at gamma-ray frequencies (ref.~\citeonline{Ardavan2021},Table~1 and Section~5.4).  It is expected, therefore, that the class of progenitors of this non-spherically decaying radiation would include, {\it inter alia}, the gamma-ray pulsars. 

To see whether this expectation is supported by the observational data on gamma-ray pulsars~\cite{Abdo2013}, we analyse these data here on the basis of the fact that, in a statistical context, luminosities and distances of the sources of a given type of radiation represent two {\it independent} random variables.  A generalization of the nonparametric rank methods for testing the independence of two random variables~\cite{Hajek} to cases of truncated data, such as the flux-limited data on gamma-ray pulsars, is the method developed by Efron and Petrosian~\cite{EF1992,EF1994}: a method that has been widely used in astrophysical contexts (see e.g.\ ref.~\citeonline{Bryant} and the references therein).   

From the raw data on fluxes $S$ and distances $D$ of gamma-ray pulsars in the second {\it FERMI} catalogue\footnote{\url{http://fermi.gsfc.nasa.gov/ssc/data/access/lat/2nd_PSR_catalog/}} and the candidate decay rates of flux density with distance ($S\propto D^{-\alpha}$ for various values of $\alpha$) we compile a collection of data sets on the prospective luminosities of these pulsars (Section~\ref{sec:data}).  We then test the hypothesis of independence of luminosity and distance by evaluating the Efron--Petrosian statistic (Section~\ref{subsec:statistic}) for the data sets on prospective luminosities and pulsar distances with three different choices of the flux threshold (i.e.\ the truncation boundary below which the data set on $S$ may be regarded as incomplete).  The resulting values of the Efron--Petrosian statistic for differing values of $\alpha$ in each case determine the significance level at which the hypothesis of independence of luminosity and distance (and hence a given value of $\alpha$) can be rejected (Section~\ref{subsec:results}).  We also assess the effects of random and (if any) systematic errors in the estimates of distance and flux on the test results by means of a Monte Carlo simulation (Section~\ref{subsec:errors}). 

\section{Observational data}
\label{sec:data}
 
\begin{figure*}
\centerline{\includegraphics[width=17cm]{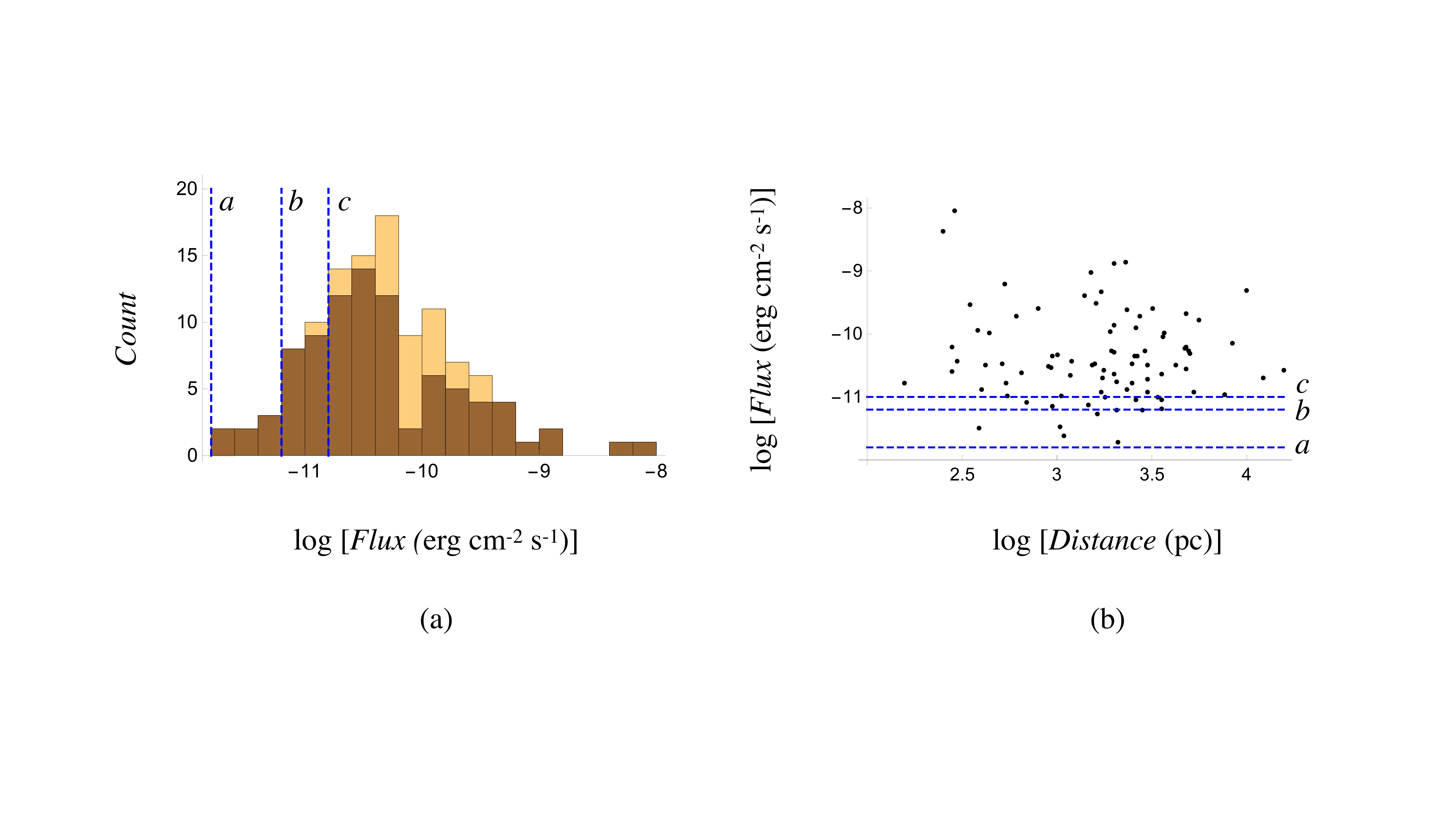}}
\caption{(a)  Histogram of the $114$ gamma-ray pulsars whose fluxes are known.   The bins in darker brown contain the $88$ pulsars for which both fluxes and distances are known.  The broken lines $a$, $b$ and $c$ each designate a flux threshold below which the plotted data set may be incomplete.  Note that the heights of the bins preceding the peak of this histogram change most sharply across the designated lines $b$ and $c$.  (b)  Distribution of logarithm of flux versus logarithm of distance for the $88$ gamma-ray pulsars in the darker brown bins.  The broken lines $a$, $b$ and $c$ designate the same flux thresholds as those shown in part (a).}
\label{LF1}
\end{figure*}

The second {\it FERMI} catalogue lists the observational data on $117$ gamma-ray pulsars of which $3$ have no flux estimates.  Histogram of the $114$ pulsars whose gamma-ray fluxes are known is shown in Fig.~\ref{LF1}a.  Of these, there are $26$ pulsars with no distance estimates.  Logarithm of the gamma-ray flux (in units of erg cm$^{-2}$ s$^{-1}$) of each of the remaining $88$ pulsars is plotted versus the logarithm of its distance (in units of pc) in Fig.~\ref{LF1}b.  The broken lines $a$, $b$ and $c$ in these figures each designate a flux threshold below which the plotted data set may be incomplete.

The isotropic gamma-ray luminosity of each pulsar is given, in terms of its flux density $S$ and its distance $D$, by 
\begin{equation}
L=4\pi \ell^2(D/\ell)^\alpha S,
\label{E1}
\end{equation}
where $\alpha=2$ if $S$ diminishes with distance as predicted by the inverse-square law and $\ell$ is a constant with the dimension of length whose value only affects the scale of $L$.  The corresponding distribution of the logarithm of $L$ (in units of erg s$^{-1}$) versus logarithm of $D$ (in units of pc) for the data set shown in Fig.~\ref{LF1}b is plotted in Fig.~\ref{LF2}a for $\alpha=2$.  The solid line in Fig.~\ref{LF2}a corresponds to the detection limit $a$ on the value of the flux density $S$ (see Fig.~\ref{LF1}).

\section{Testing the hypothesis that the data sets on luminosity and distance are independent}
\label{sec:test}
\subsection{The Efron--Petrosian statistic}
\label{subsec:statistic}

In this section we only outline the procedure by which the Efron--Petrosian statistic is calculated for two given data sets; lucid expositions of the theoretical basis of this procedure can be found in refs.~\citeonline{Maloney1999} and~\citeonline{Petrosian2002}. 

\begin{figure*}
\centerline{\includegraphics[width=18cm]{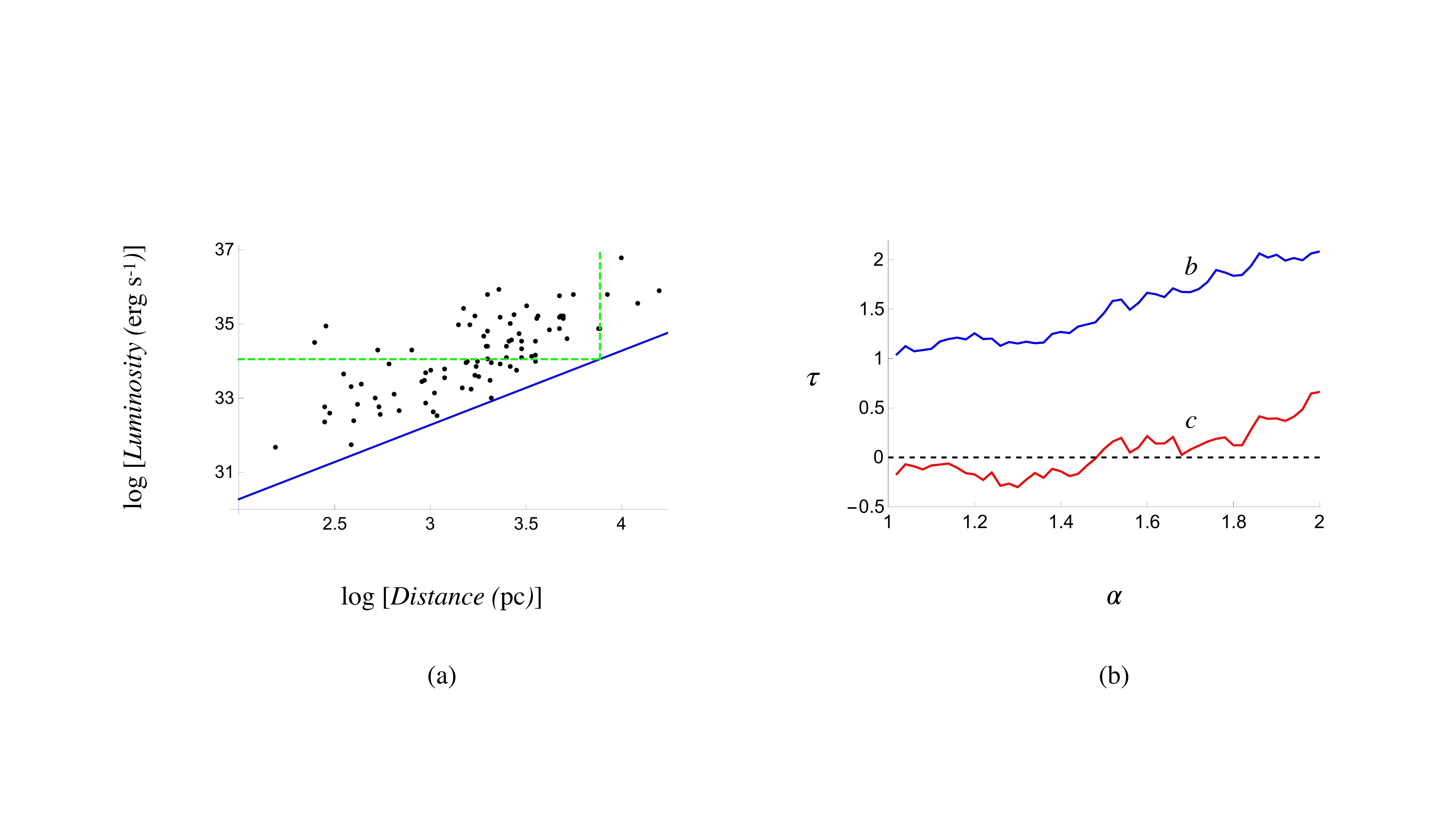}}
\caption{(a)  The luminosity-distance data set that follows from the flux-distance data set of Fig.~\ref{LF1}b and equation (\ref{E1}) for $\alpha=2$.  The solid line (in blue) is the image $L=L_{\rm th}(D)$ of the detection threshold that is marked by the dashed lines labelled $a$ in Figs.~\ref{LF1}a and \ref{LF1}b.  Those elements of this data set that lie within (and on the boundary of) the rectangular area bounded by the vertical axis and the vertical and horizontal dashed lines (in green) comprise the set comparable to the element $(3.89, 34.05)$ on the vertical dashed line. (b)  The Efron--Petrosian statistic $\tau$ versus the exponent $\alpha$ that specifies the rate of decay, $S\propto D^{-\alpha}$, of flux density with distance.  The blue and red curves in this figure respectively show the results for the flux thresholds indicated by the broken lines $b$ and $c$ in Fig.~\ref{LF1}.  The corresponding curve for threshold $a$ lies above the blue curve outside the frame of this figure.  Note that the red curve crosses the line $\tau=0$ at $\alpha=1.49$.} 
\label{LF2}
\end{figure*}

If we let $S_{\rm th}$ stand for the threshold value of flux density, then the corresponding truncation boundary for the values of luminosity (e.g.\ that shown as a solid line in Fig.~\ref{LF2}a) is given, according to equation~(\ref{E1}), by $L=L_{\rm th}(D)$ with
\begin{eqnarray}
\log L_{\rm th}&=&\log[4\pi(3.085\times10^{18})^2\ell^{2-\alpha} S_{\rm th}]\nonumber\\*
&&+\alpha\log D,
\label{E2}
\end{eqnarray}
in which the numerical factor $3.085\times10^{18}$ converts the units of $D$ and $\ell$ from pc to cm.  The data set on luminosity is thus regarded as complete only in the sector $\log L\ge \log L_{\rm th}$ of the $(\log D,\log L)$ plane.

Next, the set {\it comparable} to any given element $(\log D_i,\log L_i)$ of the bivariate distance-luminosity data set (such as the data set plotted in Fig.~\ref{LF2}a) is defined to comprise all those elements for which
\begin{equation}
\log D\le \log D_i,\qquad i=1,\cdots n,
\label{E3}
\end{equation}  
and
\begin{eqnarray}
\log L&\ge&\log[4\pi(3.085\times10^{18})^2\ell^{2-\alpha} S_{\rm th}]\nonumber\\*
&&+\alpha\log D_i,
\label{E4}
\end{eqnarray}
where $n$ is the number of elements in the part of the data set that is not excluded by the chosen flux threshold.  For instance, the set comparable to the data point $(3.89, 34.05)$ in Fig.~\ref{LF2}a consists of the elements of the data set that lie within (and on the boundary of) the rectangular region delineated by the broken lines coloured green in this figure.  We denote the number of elements in the set comparable to $(\log D_i,\log L_i)$ by $N_i$.

To determine the {\it rank} ($1\le R_i\le N_i$) of the element $(\log D_i,\log L_i)$, we now order the $N_i$ elements of its comparable set by the ascending values of their coordinates $\log L_i$ and equate $R_i$ to the position at which $(\log D_i,\log L_i)$ appears in the resulting ordered list.  Coordinates of the elements of the bivariate data set $(\log D,\log L)$ are in the present case all distinct.  So, a rank $R_i$ can be assigned to every element of this set unambiguously.  

The Efron--Petrosian {\it statistic} is given by
\begin{equation}
\tau=\frac{\sum_{i=1}^n(R_i-E_i)}{\sqrt{\sum_{i=1}^nV_i}},
\label{E5}
\end{equation}
in which
\begin{equation}
E_i=\textstyle{\frac{1}{2}}(N_i+1),\qquad V_i=\textstyle\frac{1}{12}(N_i^2-1).
\label{E6}
\end{equation}
Note that the value of $\tau$ is independent of that of the scale factor $\ell$ that appears in equation~(\ref{E1}).  

The hypothesis of independence is rejected when the value of $\tau$ falls in one of the tails of a Gaussian distribution whose mean is $0$ and whose variance is $1$.  More precisely, the hypothesis of independence is rejected when the value of
\begin{eqnarray}
p&=&(2/\pi)^{1/2}\int_{\vert\tau\vert}^\infty\exp(-x^2/2){\rm d}x\nonumber\\*
&=&{\rm erfc}(\vert\tau\vert/\sqrt{2})
\label{E7}
\end{eqnarray}
is smaller than an adopted significance level between $0$ and $1$, where ${\rm erfc}$ denotes the complementary error function.  If $\tau$ equals zero, for instance, then $p$ would assume the value one and the hypothesis in question cannot be rejected at any significance level.

\subsection{Test results}
\label{subsec:results}

In this section, we evaluate the expression in equation~(\ref{E5}) for the Efron--Petrosian statistic $\tau$, as a function of the exponent $\alpha$ that appears in the expression for luminosity in equation (\ref{E1}), for three different flux thresholds: the detection threshold $\log S_{\rm th}=-11.8$ and the thresholds $\log S_{\rm th}=-11.2$ and $\log S_{\rm th}=-10.8$ across which the bin heights of the histogram in Fig.~\ref{LF1}a change most sharply.  These three flux thresholds are marked in Fig.~\ref{LF1} by the dashed lines labelled $a$, $b$ and $c$, respectively.  The number of elements of the present  data set that are excluded by the thresholds $a$, $b$ and $c$ are $0$, $7$ and $24$, respectively.  The dependence of $\tau$ on $\alpha$ for thresholds $b$ and $c$ are respectively depicted by the blue and the red curves in Fig.~\ref{LF2}b. The corresponding curve for threshold $a$ extends over too high a range of values of $\tau$ to fall within the frame of Fig.~\ref{LF2}b.

For threshold $a$, the resulting values of $\tau$ ($\tau=5.103$ when $\alpha=2$ and $\tau=4.455$ when $\alpha=3/2$) imply that the hypothesis of independence of luminosity and distance can be rejected at significance levels as low as $5\times10^{-6}$ when $\alpha=2$ and $10^{-5}$ when $\alpha=3/2$.  This confirms that, as emphasized by Bryant et al.~\cite{Bryant} in the context of gamma-ray bursts, the value of flux below which the data are incomplete lies closer to the peak of the histogram in Fig.~\ref{LF1}a than does the detection threshold $a$.

For threshold $b$, the resulting values of $\tau$ ($\tau=2.085$ when $\alpha=2$ and $\tau=1.461$ when $\alpha=3/2$) imply the $p$ values $0.0371$ in the case of $\alpha=2$ and $0.144$ in the case of $\alpha=3/2$.  Hence, there is a wide range of significance levels at which the present hypothesis of independence can be rejected in the case of $\alpha=2$ but not in the case of $\alpha=3/2$.

What distinguishes threshold $c$ from the other two is that there exists a value of $\alpha$ in its case ($\alpha=1.49$) at which the graph of $\tau$ versus $\alpha$ over the interval $1\le\alpha\le2$ crosses the horizontal axis $\tau=0$ (see Fig.~\ref{LF2}b).  The values of $\tau$ for $\alpha=2$ and $\alpha=3/2$ are respectively given by $0.664$ and $0.085$ for threshold $c$, so that the corresponding values of $p$ are $p=0.507$ and $p=0.932$, respectively.  This means that while the hypothesis of independence of luminosity and distance for $\alpha=3/2$ cannot be rejected even at a $93\%$ significance level, this hypothesis is rejected at all significance levels exceeding $51\%$ for $\alpha=2$.

\begin{figure*}
\centerline{\includegraphics[width=17cm]{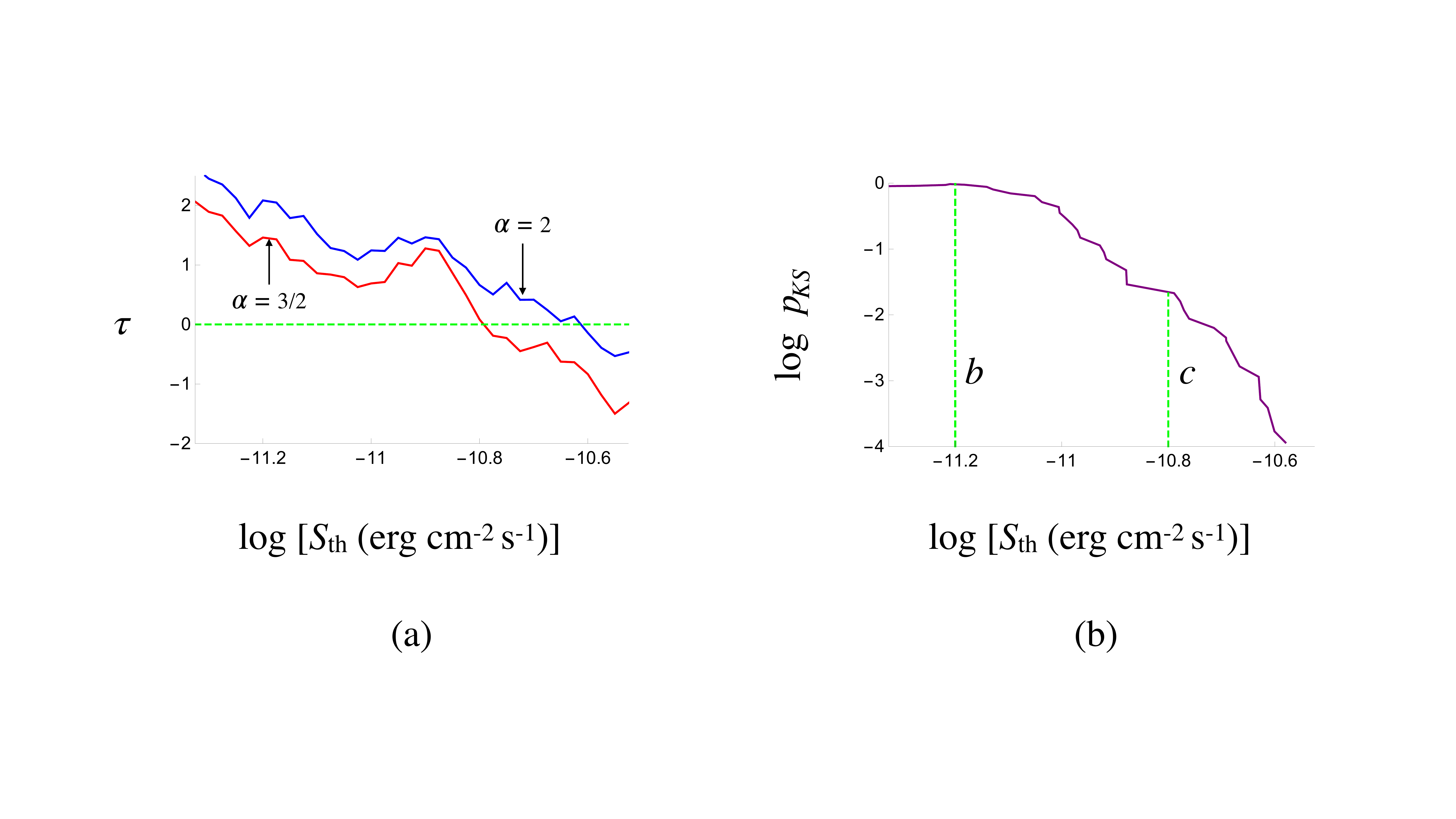}}
\caption{(a)  The Efron--Petrosian statistic $\tau$ versus the logarithm of the flux threshold $S_{\rm th}$ for $\alpha=3/2$ (the red curve) and $\alpha=2$ (the blue curve).  Note that $\tau$ vanishes when $\log S_{\rm th}=-10.79$ for $\alpha=3/2$ and when $\log S_{\rm th}=-10.61$ for $\alpha=2$.  (b) The Kolmogorov--Smirnov statistic $p_{KS}$ for testing whether the $114$-element data set on fluxes shown in Fig.~\ref{LF1}a and the truncated versions of the 88-element data set shown in Fig.~\ref{LF1}b (from which the elements with fluxes $S\le S_{\rm th}$ are eliminated) are drawn from the same distribution.  Note that while $p_{SK}$ has the values $0.97$ and $0.021$ for the thresholds $b$ and $c$, respectively, it is smaller than $1.7\times10^{-4}$ for the value of $S_{\rm th}$ at which the blue curve in part (a) crosses the line $\tau=0$.}
\label{LF3}
\end{figure*}

Figure~\ref{LF3}a shows the dependence of $\tau$ on the flux threshold $S_{\rm th}$ for both $\alpha=3/2$ and $\alpha=2$.  There is a threshold closer to the peak of the histogram in Fig.~\ref{LF1}a than $c$ for which $\tau$ vanishes when $\alpha=2$.  However, this threshold ($\log S_{\rm th}=-10.61$) would result in a truncated data set consisting of only $53$ elements that could no longer be regarded as equivalent to the original $114$-element data set depicted in Fig.~\ref{LF1}a.   According to the Kolmogorov--Smirnov test, the probability $p_{KS}$ that the resulting $53$-element data set and the original $114$-element data set are drawn from the same distribution is as low as $1.7\times10^{-4}$ (see Fig.~\ref{LF3}b).

\subsection{The effect of observational errors on the test results}
\label{subsec:errors}

Even if present, a purely systematic error in the estimates of distance and/or flux would not alter the results reported in Section~\ref{subsec:results}: the value of the Efron--Petrosian statistic does not change if we make monotonically increasing transformations on the values of distance and/or flux~\cite{EF1992}.  For instance, the dependence of $\tau$ on $\alpha$ (depicted by the curves in Fig.~\ref{LF2}b) remains exactly the same if the distances in the data set shown in Fig.~\ref{LF1}b are all multiplied by a positive factor.  In this section we perform a Monte Carlo simulation with $10^3$ random samplings to assess the effect on the test results of (the known) random errors in the estimates of distance and flux. 

Distances and fluxes are listed in the second {\it FERMI} catalogue each with a positive ($\sigma_+$) and a negative ($\sigma_-$) uncertainty (see ref.~\citeonline{Abdo2013}).  The distribution of the error in the value $\mu$ of each one of the listed variables can accordingly be modelled by an asymmetric, Gaussian probability density:
\begin{eqnarray}
f(x)&= &\sqrt{\frac{2}{\pi}}\left[\sigma_++{\rm erf}\left(\frac{\mu}{\sqrt{2}\sigma_-}\right)\sigma_-\right]^{-1}\nonumber\\*
&&\times\left\{\exp\left[-\frac{(x-\mu)^2}{2\sigma_-^2}\right]{\rm H}(\mu-x)\right.\nonumber\\*
&&\left.\quad+\exp\left[-\frac{(x-\mu)^2}{2\sigma_+^2}\right]{\rm H}(x-\mu)\right\},\nonumber\\*
&&\qquad\qquad\qquad\qquad 0<x<\infty,
\label{E8}
\end{eqnarray}
where erf and H denote the error function and the Heaviside step function, respectively.  Equation~(\ref{E8}) describes half a Guassian distribution with the mean $\mu$ and the standard deviation $\sigma_-$ in $0 \le x \le \mu$ and half a Gaussian distribution with the same mean but the standard deviation $\sigma_+$ in $\mu \le x <\infty$.  In the case of the listed fluxes, and some of the listed distances, $\sigma_+$ and $\sigma_-$ have the same values.

\begin{figure*}
\centerline{\includegraphics[width=18cm]{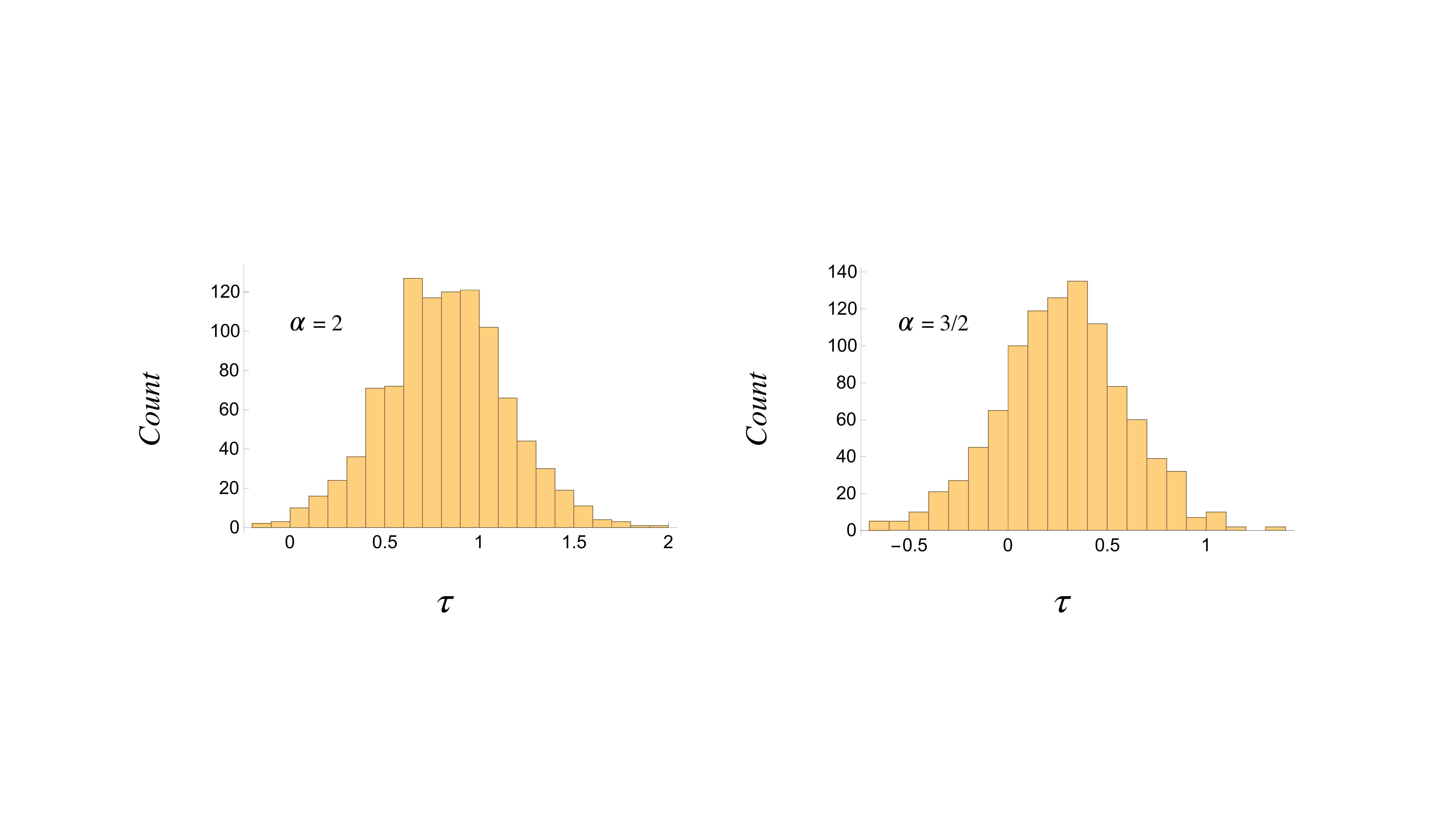}}
\caption{Histograms of the distributions of the Efron--Petrosian statistic $\tau$ for $\alpha=2$ and $\alpha=3/2$ in the case of threshold $c$.  The above distributions (obtained by the Monte Carlo simulation described in Section~\ref{subsec:errors}) show that the uncertainties in the estimates of pulsar distances and fluxes, though bringing about a spreading of the values of $\tau$, do not alter the conclusions reached on the basis of the test results depicted in Fig.~\ref{LF2}b.}
\label{LF4}
\end{figure*}

The sampling domain of the Monte Carlo method we use consists of the collection of the intervals $\mu-\sigma_- \le x  \le \mu+\sigma_+$ which enclose each of the two coordinates (whose listed value we have denoted by $\mu$) of the elements of the data set shown in Fig.~\ref{LF1}b.  In a given sampling, the values of distance and flux for each element of the data set in Fig.~\ref{LF1}b are replaced by two randomly chosen values of $x$ from the probability distribution $f(x)$ over the interval $\mu-\sigma_- \le x \le \mu+\sigma_+$, where $\mu$ and $\sigma_\pm$ are the parameters of the relevant coordinate (either distance or flux) of the data point in question as they appear in the second {\it FERMI} catalogue.  The modified data set thus obtained in a given sampling is then used, in conjunction with a choice of the flux threshold $S_{\rm th}$, to calculate the Efron--Petrosian statistic $\tau$ for both $\alpha=2$ and $\alpha=3/2$.  Repeating this procedure $10^3$ times and aggregating the two resulting sets of values of $\tau$ for differing values of $\alpha$, we arrive at the distributions shown in Fig.~\ref{LF4} in the case of threshold $c$.  (The outcome and implications of the corresponding simulation for threshold $b$ turn out to be similar to those of threshold $c$.)

The mean and the standard deviation of the $\tau$-distribution for $\alpha=2$ in Fig.~\ref{LF4} have the values $0.818$ and $0.323$, respectively, so that the uncertainties in the estimates of distance and flux result in spreading the value of the Efron--Petrosian statistic over the interval $0.495 \le \tau \le 1.141$.  According to equation~(\ref{E7}), the $p$-values corresponding to this range of values of $\tau$ occupy the interval $0.254\le p \le0.620$.  

The mean and the standard deviation of the $\tau$-distribution for $\alpha=3/2$ in Fig.~\ref{LF4} have the values $0.279$ and $0.318$, respectively.  The values of $\tau$ are in this case spread over the interval $-0.039\le\tau\le0.596$, an interval containing $\tau=0$ that corresponds to $0.551\le p\le1$.  In neither of the two cases ($\alpha=2$ and $\alpha=3/2$), therefore, are the values of the Efron--Petrosian statistic sufficiently affected by the uncertainties in the distance estimates to alter the conclusions reached on the basis of the test results reported in the preceding section: there is still a wide range of significance levels at which the hypothesis of independence of luminosity and distance can be rejected in the case of $\alpha=2$ but not in the case of $\alpha=3/2$.

\section{Discussion}
\label{sec:conclusion}

The conclusion to be drawn from the above results is that the observational data in the second {\it FERMI} catalogue are consistent with the dependence $S\propto D^{-3/2}$ of the flux densities $S$ of these pulsars on their distances $D$ at substantially higher levels of significance than they are with the dependence $S\propto D^{-2}$.  To the list of observed features of the pulsar emission (brightness temperature, polarization, spectrum and profile with microstructure and with a phase lag between the radio and gamma-ray peaks) that the analysis of the radiation by the current sheet in the magnetosphere of a neutron star has decoded (ref.\citeonline{Ardavan2021}, Section~5), we can therefore add another feature (the non-spherical decay of the gamma-ray flux) of this emission.

The violation of the inverse-square law encountered here is not incompatible with the requirements of the conservation of energy because the radiation process by which the superluminally moving current sheet in the magnetosphere of a neutron star generates the observed gamma-ray pulses is intrinsically transient.  Temporal rate of change of the energy density of the radiation generated by this process has a time-averaged value that is negative (instead of being zero as in a conventional radiation) at points where the envelopes of the wave fronts emanating from the constituent volume elements of the current sheet are cusped (ref.~\citeonline{Ardavan2021}, Fig.~1 and Section~3.2).  The difference in the fluxes of power across any two spheres centred on the star is thus balanced by the change with time of the energy contained inside the shell bounded by those spheres (ref.~\citeonline{Ardavan_JPP}, Appendix~C).

Luminosities of gamma-ray pulsars are over-estimated when the decay of their flux density $S$ is assumed to obey the inverse-square law $S\propto D^{-2}$ instead of $S\propto D^{-3/2}$ by the factor $(D/\ell)^{1/2}$ (see equation~\ref{E1}).  Value of the scale factor $\ell$ is of  the same order of magnitude as the values of the light-cylinder radii of these pulsars (ref.\citeonline{Ardavan2021}, Section~5.5).  Hence, the factor by which the luminosity of a $100$ ms gamma-ray pulsar at a distance of $2.5$ kpc is thus over-estimated is approximately $4\times10^6$.  Once this is multiplied by the ratio $\sim1/30$ of the latitudinal beam-widths of gamma-ray and radio pulsars (implied by the fraction of known pulsars that are detected in gamma-rays), we obtain a value of the order of $10^5$ for the over-estimation factor in question: a result that implies that the range of values of the correctly-estimated luminosities of gamma-ray pulsars is no different from that of the luminosities of radio pulsars.

\section*{Data availability}

The data used in this paper can be found at {\url{http://fermi.gsfc.nasa.gov/ssc/data/access/lat/2nd_PSR_catalog/}}.

\end{multicols}

\end{document}